\documentclass[11pt,longnamesfirst,preprint]{aastex}
\received{2001 November~7}
\revised{2002 March~26}
\accepted{2002 March~27}
\journalid{123}{2002 July}
\paperid{201458}
\shorttitle{Extrinsic Modulation of the META Candidates}
\shortauthors{Lazio, Tarter, \& Backus}

\begin{document}

\title{Megachannel Extraterrestrial Assay Candidates: No Transmissions
	from Intrinsically Steady Sources}
\author{T.~Joseph~W.~Lazio}
\affil{Alexandria, VA  22310}
\email{jlazio@patriot.net}

\and 

\author{Jill Tarter and Peter R.~Backus}
\affil{The SETI Institute, 2035 Landings Dr., Mountain View, CA  94043}
\email{tarter@seti.org}
\email{peter@seti.org}

%\and 
%\author{\textit{James~M.~Cordes}}

\begin{abstract}
This paper reports new, more sensitive observations of nine of the
eleven extrastatistical signals in the Megachannel Extraterrestrial
Assay (META).  These extrastatistical signals had all of the expected
characteristics of a transmission from an extraterrestrial
transmitter, except that they did not repeat.  \citeauthor{cls97}
showed that this lack of repeatability could be explained by the high
detection thresholds used in the reobservations of these candidates
combined with interstellar scintillation of intrinsically steady
sources.  We use the \citeauthor{cls97} methodology, correcting an
error in the original presentation, and our new observations to rule
out this scintillation hypothesis at a confidence level of at least
97.8\% (for the case of an intrinsically weak source) to a level in
excess of~99\% (if the source strengths are comparable to that favored
by \citeauthor{cls97}).  We also demonstrate that gravitational
microlensing cannot account for the initial detection of these
candidate signals nor is gravitational lensing likely to play a role
in future SETI programs.  We conclude that the META candidates do not
reflect a large population of powerful, strong beacons.
\end{abstract}

\keywords{extraterrestrial intelligence --- gravitational lensing -- scattering}

\section{Introduction}\label{sec:intro}

Various searches for extraterrestrial intelligence (SETI) have found
signals having all of the expected characteristics of hypothesized ET
transmitters, except one.  The candidate signals are narrowband, often
are in or near a ``magic frequency'' in a ``special'' reference frame
(e.g., the Galactic barycenter), do not match the characteristics of
known, interfering signals of terrestrial or solar system origin, and
yet, do not repeat when follow-up observations of the relevant sky
positions are performed.  Among the relevant programs are the
Megachannel Extraterrestrial Assay \citep[META,][ hereinafter \citeauthor{hs93}]{hs93}, which found
several dozen such signals at~1420 and~2840~MHz, and the old Ohio
State SETI program \citep{d85}, which found the ``WOW!'' signal
at~1420~MHz.  (The SERENDIP programs [\citealt*{bwd94}] define candidate
signals as those that have been detected multiple times at the same
sky position, but only modest efforts have been made to analyze these
candidates further and it is not clear yet if these candidates can be
eliminated as intrinsically steady sources in the way that the META
candidates are below.)

By contrast with such an idealized type of signal, it is easy to
imagine that received signals might depart significantly from this
idealization, particularly with regard to duration and amplitude
stability.  \citefullauthor{cls97}~(\citeyear{cls97}, hereinafter
\citeauthor{cls97}) listed
(possibly incompletely) reasons for a received signal to appear
intermittent:
\begin{enumerate}
\item Noise in the receiver electronics, e.g., thermal noise, cosmic
ray induced events, and hardware failures;
\item Radio frequency interference (RFI) whose origin is terrestrial,
from Earth orbit, or from interplanetary spacecraft;
\item Natural, extrinsic modulation of a (unknown) class of narrowband
astrophysical sources such as that caused by interstellar (radio)
scintillation (ISS) or gravitational lensing;
\item Natural, extrinsic modulation of ETI sources; and
\item Intrinsic intermittency at the source of an ETI signal due to
natural causes, such as planetary rotation or the nature of the
transmission (e.g., planetary radar) or, for deliberate reasons, to
frustrate detection and decryption by non-target civilizations.
\end{enumerate}

They then proceeded to analyze the impact of Cause~4 upon SETI
surveys, with particular emphasis on ISS, motivated by the following
considerations.  First, it seems to be the simplest mechanism for
producing transient signals from otherwise intrinsically steady
signals that is amenable to testing.  Second, observations of pulsars
and active galactic nuclei have demonstrated that ISS is important at
centimeter wavelengths \citep{r90} that are commonly used in searches
for \hbox{ETI} \citep{o73}.

\citeauthor{cls97} were able to show that ISS \emph{by itself} is
sufficient to explain the lack of confirmation of any ETI candidate
signals in surveys conducted to date.  They also predicted the
sensitivities required to rule out the scintillation modulation
hypothesis.

This paper reports new observations of nine of the eleven META candidates,
observations with the requisite sensitivity to allow ISS to be ruled
out.  In \S\ref{sec:meta} we summarize the extrastatistical candidates
from the META and what the ISS hypothesis requires about their signal
strengths, in \S\ref{sec:observe} we describe the additional
observations we have conducted.  We demonstrate that ISS cannot be
responsible for modulating an intrinsically steady signal to produce
the META candidates in \S\ref{sec:iss} and that gravitational
microlensing cannot do the same in \S\ref{sec:micro}.  We summarize
our conclusions in \S\ref{sec:conclude}.

\section{The Extrastatistical META Candidates}\label{sec:meta}

In this section we summarize briefly the relevant properties of the
META extrastatistical signals.  For full details of META and the various
signals detected, see \citeauthor{hs93}.

META conducted five surveys of the northern sky, three at the
frequency 1420~MHz and two at~2840~MHz, between~1986 and~1991.  At
each sky position observed, two polarizations were searched in each of
three reference frames---the Local Standard of Rest, the Galactic
barycenter, and the cosmic microwave background.  The vast majority of
the signals analyzed were consistent with an exponential distribution
as expected for noise from a Fourier transform spectrometer.

Of the roughly $6 \times 10^{13}$ observations analyzed during the
five-year META, only eleven could not be explained as due to either
noise or processor failure, e.g., due to cosmic-ray hits on the
electronics.  It was these eleven extrastatistical detections, four
at~1420~MHz and seven at~2840~MHz, that \citeauthor{hs93} identified
as being candidates for detections of ETI transmitters.

All eleven candidates were narrowband, with the broadest being only
four spectrometer channels wide; each channel was 0.05~Hz.  Some
candidates were unresolved.  During the survey, the processing
software could halt the survey observations and acquire reobservations
on ``interesting'' signals within~40~s of detection.  (The eleven
candidates were identified only after the conclusion of META so far
more than just the eleven candidates were reobserved in this manner.)
In no case was a candidate reobserved during the immediate
reobservations.  Furthermore, various followup observations were
attempted over a five-year period, and no candidate was detected
again.
(Intense scrutiny of the WOW sky position has also
failed to redetect a signal, \citealt{g94,gm01})

With the exception of the lack of reobservation, the candidates had
all of the characteristics expected of ETI transmitters:  Narrowband
signals in a celestial reference frame.  The eleven extrastatistical
candidates also all had low Galactic latitudes, consistent with that
expected for a Galactic population.  \citeauthor{hs93} also considered
the possibility that these candidates represent signals from a
previously unrecognized class of natural sources.  Throughout the rest
of this paper we shall discuss the candidates as if they are ETI
transmitters, but our comments will be equally applicable if they
represent a natural class of sources.

\section{Observations}\label{sec:observe}

We observed nine of the eleven META candidates during the interval
1997--1998 as part of Project Phoenix \citep{c99}.
Table~\ref{tab:log} summarizes the observing log.  We also report the
candidate positions in a common and conventional equinox (J2000).

\begin{deluxetable}{cccc}
\tablecaption{META Followup Observational Log\label{tab:log}}
\tablewidth{0pc}
\tablehead{
 \colhead{Right Ascension} & \colhead{Declination} & \colhead{Epoch} & 
 \colhead{$\eta_1$}\\
 \multicolumn{2}{c}{(J2000)}}
\startdata
\cutinhead{1420~MHz candidates}

03 06 46.87 & $+58$ 02 19.4 & 1997 December~17 & 224 \\
            &               & 1998 February~5 \\
            &               & 1998 April~28 \\
19 34 29.27 & $+47$ 31 19.7 & 1997 December~19 & 35.6 \\
            &               & 1998 April~28 \\
21 59 20.33 & $+38$ 33 40.5 & 1997 December~11 & 33.6 \\
            &               & 1998 April~28 \\
23 43 38.89 & $+08$ 33 02.2 & 1997 December~11 & 33.0 \\
            &               & 1998 April~28 \\

\\ 

\cutinhead{2840~MHz candidates}

08 00 33.13 & $-08$ 31 54.4 & \ldots\tablenotemark{a} & 747 \\
08 57 31.47 & $-15$ 47 36.7 & \ldots\tablenotemark{a} & 75.4 \\
14 18 22.30 & $+57$ 26 44.9 & 1998 March~23 & 31.8 \\ 
14 39 26.08 & $+46$ 26 55.3 & 1998 March~23 & 31.8 \\
18 22 51.59 & $-19$ 29 38.0 & 1998 March~23 & 52.8 \\
18 40 52.38 & $-23$ 14 20.9 & 1998 March~23 & 44.4 \\
20 02 17.13 & $+30$ 47 04.1 & 1998 March~23 & 33.2 \\

\enddata
\tablenotetext{a}{This candidate was not reobserved as part of this work.}
\tablecomments{$\eta_1$ is the initial detection strength of the
candidate during the META, in units of the mean META noise level.}
\end{deluxetable}

For a complete discussion of the observational methodology of Project
Phoenix, see \cite{c99}.  We shall summarize only the relevant
details.  Coordinated observations were conducted with the 43~m
(140~ft) NRAO telescope at Green Bank, WV, and a 30~m telescope in
Woodbury, \hbox{GA}.  Initial observations were conducted with the
NRAO telescope employing a filter to excise terrestrial interference.
The detection of a possible signal exceeding a specified threshold
(see below) was then transmitted to the telescope in Woodbury.
Employing a matched filter and a lower threshold, the telescope at
Woodbury re-observed the sky position.  A signal had to be detected at
both telescopes in order to be considered a genuine celestial signal.

The original META observations used a telescope with a 0\fdg5
beamwidth (FWHM) at a frequency of~1420~MHz.  In order to cover this
area fully with our smaller beam, 0\fdg3 at~1420~MHz, a grid of
positions centered on the nominal position of the META candidate was
observed.  The grid included the nominal position of the candidate and
pointings separated by 0\fdg3 north, east, south, and west of the
nominal position.

At each pointing position an integration time of~300~s was used.  The
bandwidth of the individual spectral channels was 0.68~Hz (vs.\
0.05~Hz in META).  The equivalent ($1\sigma$) sensitivity was $1.4
\times 10^{-27}$~W~m${}^{-2}$, and the initial detection threshold for
the NRAO telescope was $7\sigma$.  For comparison the typical META
sensitivity ($1\sigma$) was $5.7 \times 10^{-25}$~W~m${}^{-2}$, and
the threshold for the extrastatistical candidates described in
\S\ref{sec:meta} was $31.7\sigma$.

\citeauthor{cls97} normalized intensities to the mean noise
level~$\langle N\rangle$ in the META spectrometer.  We shall continue
to do so as well.  The choice of whether to use the META or Project
Phoenix noise level is arbitary, but because the intrinsic signal
strength of the (putative) transmitters is also normalized by $\langle
N\rangle$, we must be consistent.  Denoting the normalized intensities
by $\eta$ (as \citeauthor{cls97} did), the various observational
thresholds described above become $\eta = I/\langle N\rangle = 31.7$
for META, $\eta = 0.018$ for the NRAO telescope, and $\eta = 0.0045$
for the telescope at Woodbury.  An important quantity for our analysis
will be the dynamic range between the initial candidate detection
level~$\eta_1$ and the subsequent reobservation threshold~$\eta_T$.
For our observations, $\eta_1/\eta_T \sim 10^3$.  For reference,
$\eta_1/\eta_T \sim 2$ for the original META reobservations.

\section{Interstellar Scintillation}\label{sec:iss}

In this section we first summarize key results from
\citeauthor{cls97}.  We then use the formalism developed by
\citeauthor{cls97} to show that our new observations exclude the META
candidates from being intrinsically steady ETI transmitters.

Compact radio sources observed at centimeter wavelengths (e.g.,
pulsars) scintillate.  Sufficiently distant radio sources ($\gtrsim
0.5$~kpc) are in the saturated scintillation regime.\footnote{%
We emphasize that like \citeauthor{cls97} we assume that saturated
scintillations, in which the rms intensity modulation is 100\%, apply.
This is unlikely to be the case for the rest of Project Phoenix
observations, as most of its observations target nearby stars, so the
distance to any potential transmitters is not sufficient for the
saturated scintillation regime to obtain.  More distant sources in the
beam may be in the saturated scintillation regime, though.%
}  The signal from an intrinsically steady source of flux density~$S$
will have an observed flux density of $gS$ (in the absence of noise).
The probability density function of the ISS gain~$g$ is $p(g) =
e^{-g}U(g)$, where $U(g)$ is the Heaviside step function.  A point
discussed at length in \cite{cl91} and \citeauthor{cls97} is the
ancipital nature of this probability density function: Scintillations
can act to render an otherwise undetectable signal detectable, but
because the most probable gain is $g = 0$, ISS will more likely render
undetectable an otherwise detectable signal.

A key result of \citeauthor{cls97} was the role of noise in
\hbox{META}.  They showed that the META candidates could be explained
as rare combinations of a gain $g > 1$ \emph{and} a noise spike.  The
ISS gain has a decorrelation time scale of minutes to days, depending
upon a source's Galactic longitude and latitude and the velocities of
the source, intervening medium, and observer.  The noise in the
spectrometer decorrelated on the time scales required to compute a
fast Fourier transform ($\simeq 20$~s).  The failure of the immediate
reobservations to detect the source could be understood as due to a
reobservation threshold that was too high coupled with the noise
decorrelating; later reobservations had the additional complication
that the ISS gain had also decorrelated either partially or fully.

\citeauthor{cls97} quantified the probability of redetecting a source
with the conditional probability $P_{\mathrm{2d}}(I_T | I_1; S, \rho)$
(\citeauthor{cls97}, Appendix~B).  Here a source of intrinsic
strength~$S$ is detected initially at a level~$I_1$.  At a later time,
for which the ISS gain correlation coefficient is $\rho$ (with $0 \le
\rho \le 1$), the source is then reobserved with a reobservation
detection threshold of $I_T$.  Following \citeauthor{cls97} the
quantities can be normalized by the noise level at the time of initial
detection, $P_{\mathrm{2d}}(\eta_T | \eta_1; \zeta, \rho)$, where
$\zeta \equiv S/\langle N\rangle$.  However, the expression given  by
\citeauthor{cls97} (their equation~[B19]) is incomplete.  The correct formulation of
$P_{\mathrm{2d}}(I_T|I_1; S, \rho)$ is
\begin{equation}
P_{\mathrm{2d}}(I_T|I_1; S, \rho) 
 = \frac{\int_{I_T}^\infty\int_{I_1}^\infty dI_2^\prime dI_1^\prime\,f_{2I}{(I_1^\prime, I_2^\prime; S,\rho)}}{P_{d,\mathrm{scint}}(I_1; S)}.
\label{eqn:p2d}
\end{equation}
The joint intensity probability density function is $f_{2\eta}(\eta_1, 
\eta_2; \zeta, \rho)$ and is given by equation~(B13) of
\citeauthor{cls97}.  The signal detection probability of a
scintillating source, above a threshold~$\eta_1$, is
$P_{d,\mathrm{scint}}(\eta_1, \zeta)$ and is given by equation~(B10)
of \citeauthor{cls97}.  

\begin{figure}
\vspace*{-0.1cm}
\epsscale{0.7}
\rotatebox{-90}{\plotone{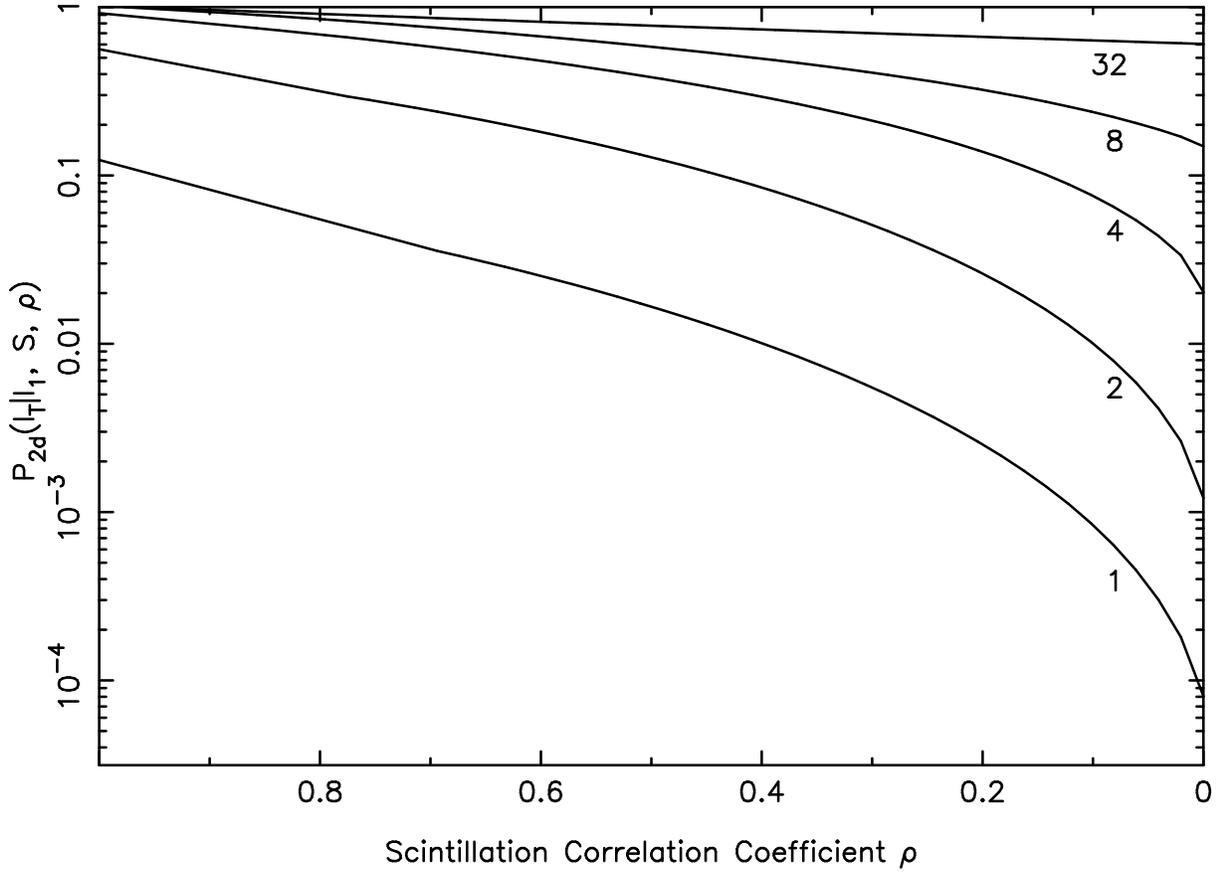}}
\caption[]{The second detection probability
$P_{\mathrm{2d}}(\eta_T|\eta_1; \zeta, \rho)$, using the corrected
expression of equation~(\ref{eqn:p2d}), for values of the
initial detection threshold, $\eta_1 \approx 32$, and redetection
threshold, $\eta_1 \approx 20$, relevant to the META observations.
The curves are labeled by the 
value of the intrinsic signal strength~$\zeta$.  This figure should
be compared to Figure~3 of \citeauthor{cls97}.}
\label{fig:p2d}
\end{figure}

For completeness Figure~\ref{fig:p2d} shows the corrected
$P_{\mathrm{2d}}(\eta_T|\eta_1; \zeta, \rho)$ for the values
of~$\eta_T$ and~$\eta_1$ relevant to the META observations and with
various values of~$\zeta$.  Figure~\ref{fig:p2d} reproduces one of the
panels of Figure~3 in \citeauthor{cls97}, but with the corrected
expression for $P_{\mathrm{2d}}(\eta_T|\eta_1; \zeta, \rho)$,
equation~(\ref{eqn:p2d}).  Comparison of Figure~\ref{fig:p2d} and
Figure~3 in \citeauthor{cls97} shows that the figures in
\citeauthor{cls97} that display $P_{\mathrm{2d}}(\eta_T|\eta_1; \zeta,
\rho)$ are qualitatively correct and largely quantitatively correct.

Scintillations impact our observations in two ways.  First, the
original motivation in conducting the observations was to use detection 
thresholds low enough to rule out the scintillation hypothesis for the 
META candidates.  Second, however, our use of the two telescopes means 
that a scintillating signal could be undetected in the reobservations
with the telescope at Woodbury.

We consider first our coincidence detection scheme, because if this
scheme is not robust against scintillations, we can draw no
conclusions about the comparisions with the META observations.  In the 
coincidence detection scheme the initial observation threshold is that 
of the NRAO telescope, $\eta_1^\prime = 0.018$, and the reobservation
threshold is that of the telescope at Woodbury, $\eta_T^\prime =
0.0045$.  The degree of correlation of the scintillations is unknown,
but we consider two possibilities to illustrate the possible range of
second detection probabilities.  

In the case $\rho = 1$, there exists no analytic expression for
$P_{2d}(\eta_T|\eta_1; \zeta, \rho=1)$, but we can develop a useful
approximation.  The bivariate intensity density function is 
\begin{equation}
f_{2\eta}(\eta_1, \eta_2; \zeta, \rho=1) \\
 = \frac{1}{1+2\zeta}\,I_0\left(\frac{\zeta\sqrt{\eta_1\eta_2}}{1+2\zeta}\right)\exp\left[-\frac{(\eta_1+\eta_2)(1+\zeta)}{1+2\zeta}\right],
\label{eqn:f2i}
\end{equation}
where $I_0(x)$ is the first order modified Bessel function.
We rewrite $P_{2d}(\eta_T|\eta_1; \zeta, \rho=1)$ as
\begin{equation}
P_{2d}(\eta_T|\eta_1; \zeta, \rho=1)
 = \frac{1-\int_0^{\eta_T} d\eta_2\,\int_0^{\eta_1} d\eta_1\,f_{2\eta}(\eta_1, \eta_2; \zeta, \rho=1)}{P_{d,\mathrm{scint}}(\eta_1;\zeta)}
\label{eqn:p2drewrite}
\end{equation}
and then, because $\eta_T^\prime$, $\eta_1^\prime \ll 1$, we expand in powers
of~$\eta_T^\prime$ and~$\eta_1^\prime$.  We find
\begin{equation}
P_{2d}(\eta_T|\eta_1; \zeta, \rho=1)
 \approx \left\{1-\frac{\eta_T\eta_1}{1+2\zeta}\left[1-\frac{(1+\zeta)(\eta_T+\eta_1)}{2(1+2\zeta)}+\frac{\zeta^2\eta_T\eta_1}{(1+2\zeta)^2}\right]\right\}e^{\eta_1/(1+\zeta)}.
\label{eqn:p2dapprox}
\end{equation}
In order to explain the META candidates as scintillating sources,
\citeauthor{cls97} also found that $\zeta \sim 3$.  For all reasonable
values of~$\zeta > 0$, the redetection proabability for the Project
Phoenix coincidence detection procedure is
$P^{(P)}_{2d}(\eta_T^\prime|\eta_1^\prime; \zeta, \rho=1) > 0.999$.

If the scintillations are uncorrelated completely, $\rho = 0$, then
$P_{\mathrm{2d}}$ can be evaluated analytically.
\begin{equation}
P_{\mathrm{2d}}(\eta_T | \eta_1; \zeta, \rho=0)
 = P_{d,\mathrm{scint}}(\eta_T; \zeta)
 = e^{-\eta_T/(1+\zeta)}
\label{eqn:pdproduct}
\end{equation}
For a source of strength comparable to that preferred by
\citeauthor{cls97}, $P^{(P)}_{\mathrm{2d}}(\eta_T^\prime|\eta_1^\prime; \zeta, \rho=0) \simeq
0.999$.  A strict \emph{lower} limit to the detection probability is
obtained by setting $\zeta = 0$ (implying that no source is present!)
for which $P_{\mathrm{2d}}^{(P)} \ge 0.996$.

Thus, the coincidence detection scheme is at least 99.6\% robust in
detecting actual scintillating sources of the source strength
estimated by \citeauthor{cls97}. Unfortunately, we can place no
reasonable limits on $\rho$.  In order for $\rho \ll 1$, the spatial
scale length of the scintillations would have to be comparable to or
smaller than the baseline between NRAO and Woodbury, $b \approx
500$~km.  At~1.4~GHz ($\lambda = 21$~cm), such a scintillation scale
length is equivalent to a scattering angle $\theta_d \sim \lambda/b
\sim 0\farcs09$.  While large, a scattering angle of this magnitude is
not unprecedented.  Various low Galactic latitude sources have
scattering diameters comparable to or larger than this value (e.g.,
\objectname[NGC]{NGC~6334B}, \citealt*{mrgb90}; \objectname[]{Cyg~X-3}, \citealt*{mmrj95}).  As the
META candidates have low Galactic latitudes as well, we can place no
constraints on~$\rho$.

In the roughly 10~yr between the META observations and the
observations we report (\S\ref{sec:observe}), scintillations would 
have decorrelated completely.  In this case the expression for
$P_{\mathrm{2d}}$ is given in equation~(\ref{eqn:pdproduct}).  For the 
META reobservations, the reobservation threshold is that of the NRAO
telescope, $\eta_T = 0.018$.  Thus, $P^{(M)}_{\mathrm{2d}}(\eta_T |
\eta_1; \zeta\sim3, \rho=0) = 0.996$ for a source strength comparable
to that estimated by \citeauthor{cls97}, and the lower limit is
$P^{(M)}_{\mathrm{2d}}(\eta_T | \eta_1; \zeta=0, \rho=0) = 0.982$.

In order to find the overall probability of detection, we multiply the
detection probability from the coincidence scheme with the redetection
probability to the original META observations, $P_{\mathrm{2d}} = P^{(P)}_{\mathrm{2d}}P^{(M)}_{\mathrm{2d}}$.  The strict lower limit
to the overall redetection probability ($\zeta = 0$) is 97.8\%, while
it is in excess of~99.5\% for a source strength comparable to that
estimated by \citeauthor{cls97}.

These confidence levels are calculated using the nominal META
sensitivity.  The actual sensitivity varied from $2.3 \times
10^{-25}$~W~m${}^{-2}$ to $4.3 \times 10^{-24}$~W~m${}^{-2}$.
\citeauthor{hs93} do not report the appropriate value for each
candidate.  Using the highest sensitivity reduces the overall
redetection probability to $P_{\mathrm{2d}}(\zeta \sim 3) \ge 98.6$\%
while using the lowest sensitivity increases the overall redetection
probability to $P_{\mathrm{2d}}(\zeta \sim 3) \ge 99.9$\%.

\citeauthor{cls97} also show the number of (uncorrelated)
reobservations required to exclude completely the possibility that the
META candidates were actual ETI transmitters.  In the case that
$\eta_T \to 0$, the number of reobservations required is 10.  Because
the maximum number of reobservations for any one candidate is only 3
(at our sensitivity level), we cannot apply this second reality
criterion.

\section{Gravitational Microlensing}\label{sec:micro}

As a second extrinsic cause of signal modulation, we consider
gravitational microlensing.  In this scenario, the continuous
transmissions of an ETI transmitter are amplified briefly by a
foreground object passing close to the line of sight of the
transmitter.  This scenario was not covered by \citeauthor{cls97}.
Gravitational microlensing has an apparent advantage over ISS in that
the amplification gain for microlensing~$A$ is $A \ge 1$.
\cite{br96}, \cite{alcocketal96}, and \cite{d99} consider the more
general problem of detecting terrestrial planets via gravitational
lensing.

Gravitational microlensing has been detected in optical monitoring
programs of stars toward the Galactic bulge, the Magellanic Clouds,
and \objectname[M]{M~31}
\citep[e.g.,][]{uskkmk94,alcocketal95,ansarietal96}.  Because
gravitational effects are achromatic, we can make use of the formalism
developed to describe these monitoring programs.  We shall assume that
neither source size nor source blending are important as both can
contribute to apparent chromatic effects \citep{hpj00}.

Given the observed signal strengths of the META candidates and the
reobservation thresholds in our new observations, the microlensing
amplifications required are $A \gtrsim 100$, somewhat smaller than the
ratio $\eta_1/\eta_T$.    Although we have not conducted extensive tests,
a la \citeauthor{cls97}, an important conclusion of \citeauthor{cls97}
continues to hold: The noise in the META spectrometer would have
played a key role in the initial detections, if these are otherwise
constant signals.  The decorrelation time scale of the noise,
approximately 20~s, was a crucial feature of the signal model
developed by \citeauthor{cls97}.  This short decorrelation time
explained how the immediate reobservations conducted during META were
unable to detect the signal.  Like ISS, the decorrelation time of
gravitational microlensing, $\Delta t_{\mathrm{gl}} > 2$~hr
\citep{p86}, is much longer than the time it took for the immediate
reobservations to occur.  Thus, gravitational microlensing need not
account for the full dynamic range $\eta_1/\eta_T$.  Amplifications
this large have been termed extreme gravitational lensing events
(EGLEs) or extreme magnification events (EMEs) \citep{p95,wt96,g97}.

\cite{p86} determined the probability for the gravitational
microlensing amplification to exceed a fiducial value~$A_0$,
$P_{\mathrm{gl}}(A>A_0)$.  For $A_0 \gg 1$ this function can be
approximated as $P_{\mathrm{gl}}(A>A_0) \simeq \tau A_0^{-2}$, where
$\tau$ is the optical depth to microlensing.  The exact values for
$\tau$ depend upon the details of the population of lensing objects.
Existing estimates for the microlensing optical depth toward the inner
Galaxy are sufficient though to place severe constraints on the META
candidates.  The optical depth toward the inner Galaxy is $\tau
\sim 10^{-6}$ \citep{kp94,uskkmk94,alcocketal95}.  This estimate
depends upon such factors as the mass function and Galactic
distribution of the lenses \citep{g99}.  In particular, the value can
be lower by factors of several, depending upon the Galactic
coordinates of the line of sight.  For the purposes of this analysis,
we ignore these differences.  Including these effects would make our
conclusions more robust.

In order to explain the META candidates, we require $A_0 \sim 100$ so
that $P_{\mathrm{gl}}(A>A_0) \sim 10^{-10}$ toward the inner Galaxy.
On average, we expect that in order to obtain an amplification as
large as $A_0$, the number of \emph{background} objects, i.e., ETI
transmitters, must be $NP_{\mathrm{gl}} \sim 1$.  The relevant
quantity here is the number of background objects as the number of
lenses has already been incorporated through the factor of $\tau$ in
$P_{\mathrm{gl}}$.

We therefore require $N \gtrsim 10^{10}$ in order to explain the META
candidates as being real signals modulated by gravitational
microlensing.  \citeauthor{hs93} estimate that if all of the META
candidates represent real signals, the Galactic population of such
transmitters is only roughly $2 \times 10^6$.  The observed number of
candidates is insufficient, by orders of magnitude, to allow them to
be explained as gravitational microlensed transmitters.

We can extend this analysis to future surveys as well.  Suppose that
we consider lower amplification values.  The number of transmitters
required within the inner Galaxy must still be $N \sim \tau^{-1} \sim
10^6$ for gravitational microlensing to be an important factor.  This
estimate for the gravitational lensing optical depth is for background
sources located in the Galactic bulge.  We assume that these sources
are located within a Galactocentric radius $R = 1$~kpc.  Then the
surface density of such transmitters must be
$0.3\,\mathrm{pc}^{-2}\,(R/1\,\mathrm{kpc})^{-2}$, assuming a
disk-like distribution of transmitters.  We have assumed a disk-like
distribution of transmitters based on the apparent concentration of
META candidates to the Galactic plane.  Extending this distribution
into the Galactic disk, the average distance between the ETI
transmitters would be roughly $0.6\,\mathrm{pc}\,(R/1\,\mathrm{kpc})$.
Given that programs such as the SETI Institute's Project Phoenix have
not detected any transmissions from nearby stars, this surface density
is far higher than the actual surface density.  We conclude that
gravitational microlensing is unlikely to play an important role in
modulating the signals from ETI transmitters within the Galaxy, unless
there are populations of disk and inner Galaxy ETI transmitters whose
densities differ significantly.

\section{Conclusions}\label{sec:conclude}

We have reported additional, more sensitive observations of nine of
the eleven extrastatistical candidates from the Megachannel
Extraterrestrial Assay (Table~\ref{tab:log}).  We have used these
observations to evaluate the most simple hypothesis amenable to
testing (\citeauthor{cls97}): That the extrastatistical candidates
were intrinsically steady sources, either extraterrestrial
transmitters or an as-yet unknown population of narrowband natural
sources, whose intensities were amplified by propagation effects,
either interstellar scintillations (\S\ref{sec:iss}) or gravitational
microlensing (\S\ref{sec:micro}).

Using the formalism developed in \citeauthor{cls97} (with a correction
for the second detection probability, equation~\ref{eqn:p2d}), we have
shown that these more sensitive limits exclude the scintillation
hypothesis: The META candidates do not in general represent a
population of scintillating sources.  This hypothesis can be excluded
at a confidence level of at least 97.8\% (for the case of an
intrinsically weak source) to a level in excess of~99\% (if the source
strengths are comparable to that favored by \citeauthor{cls97}).
Because two of the META candidates were not reobserved with this
sensitivity, we cannot rule out the possibility that one or both of
them represent ETI transmitters.

We have also shown that the number of transmitters in the inner Galaxy
would have to approach $10^{10}$ if gravitational microlensing is to
be invoked to account for the META candidates.  Current constraints on
the number of transmitters in the Galaxy are orders of magnitude
smaller than this number.  Furthermore, existing surveys already
suggest that the Galactic population of ETI transmitters is
sufficiently small so that gravitational microlensing is unlikely to
play a role in future surveys.

\citeauthor{hs93} derived various limits on various populations of
transmitters.  These limits can be revisited in light of our results
(see Figure~10 of \citeauthor{hs93}): There is no more than 1
Kardashev Type~II civilization that has constructed an isotropic
beacon near the \ion{H}{1} line or its second harmonic within the
Galaxy or nearest $10^3$ galaxies.  There are no more than $10^4$
Kardashev Type~I civilizations broadcasting isotropically within the
Galaxy and no more than 1 such civilization with a directed beacon
transmitter having 30~dB or more gain anywhere within the Galaxy.
More stringent limits on weaker transmissions will have to await
future surveys.

\acknowledgements
We thank J.~Cordes for helpful discussions, Jaem for assistance with
equation~(\ref{eqn:f2i}), and the referee for
suggestions that improved the presentation of our results.

\end{document}